\documentclass[10pt]{article}
\usepackage{cite}

\usepackage[english]{babel}
\usepackage[letterpaper,top=2cm,bottom=2cm,left=3cm,right=3cm,marginparwidth=1.75cm]{geometry}
\usepackage{amsmath}
\usepackage{amssymb}
\usepackage{aas_macros_prc}
\usepackage{graphicx}
\usepackage{color}
\usepackage[colorlinks=true, allcolors=blue]{hyperref}
\usepackage[capitalize]{cleveref}
\usepackage{siunitx}
\usepackage{csquotes}
\usepackage[table]{xcolor}
\usepackage[font=footnotesize,labelfont=bf]{caption}
\usepackage{wrapfig}
\usepackage{soul}

\definecolor{DeepSpace}{RGB}{0,50,71} \definecolor{White}{rgb}{1,1,1}

\DeclareSIPrePower{\root}{1/2}
\DeclareSIUnit{\rthz}{\sqrt{\unit{Hz}}}
\DeclareSIUnit{\femtog}{femto\text{-}g}
\DeclareSIUnit{\rtHz}{\rthz}
\DeclareSIUnit{\one}{1}
\DeclareSIUnit[per-mode=power]{\perrtHz}{\per\root\Hz}

\newcommand{\sol}{\ensuremath{\odot}}
\newcommand{\Msolar}{\ensuremath{\mathrm{M}_\sol}}

\begin{document}

\title{Science of the LISA mission: A Summary for the European Strategy for Particle Physics}

\date{\vspace{-5ex}}
\maketitle

\begin{center}
Chiara Caprini$^{a,b,*}$, Anna Heffernan$^{c}$ for \textbf{\href{https://www.cosmos.esa.int/web/lisa/lisa-science-team}{The LISA Science Team}} \\
\vspace*{0.5cm}
Additional authors:\\ 

Richard Brito$^{d}$, Gabriele Franciolini$^{a}$, Germano Nardini$^{e}$, Nicola Tamanini$^{f}$, Dani\`ele Steer$^{g}$
\end{center}

%\begin{center}

%\author[a,b]{Chiara Caprini,}
%\author[c]{Anna Heffernan} 
%\author[]{for \textbf{\href{https://www.cosmos.esa.int/web/lisa/lisa-science-team}{The LISA Science Team}}} 
%\author[d]{Richard Brito,}
%\author[a]{Gabriele Franciolini,} 
%\author[e]{Germano Nardini,}
%\author[f]{Nicola Tamanini,}
%\author[g]{Dani\`ele Steer}
%\end{center}
%\vspace*{1cm}

\noindent\textit{\small{$^a$ {CERN, Theoretical Physics Department, Esplanade des Particules 1, Geneva 1211, Switzerland}\\
$^b$ {D\'epartement de Physique Th\'eorique, Universit\'e de Gen\`eve,
CH-1211 Gen\`eve, Switzerland}\\
$^c$ {Departament de Física, Universitat de les Illes Balears, IAC3 – IEEC, Crta. Valldemossa km 7.5, E-07122 Palma, Spain}\\
$^d$ {CENTRA, Departamento de Física, Instituto Superior Técnico – IST Universidade de Lisboa – UL, Avenida Rovisco Pais 1, 1049-001 Lisboa, Portugal}\\
$^e$ {Department of Mathematics and Physics, University of Stavanger, NO-4036 Stavanger, Norway}\\
$^f$ {Laboratoire des 2 Infinis - Toulouse (L2IT-IN2P3), Université de Toulouse, CNRS, F-31062 Toulouse Cedex 9, France}\\
$^g$ {Laboratoire de Physique de l’\'Ecole Normale Sup\'erieure, ENS, CNRS, Universit\'e PSL, Sorbonne Universit\'e, Universit\'e Paris Cit\'e, F-75005 Paris, France}}}

\vspace*{1cm}

\begin{abstract}
The LISA mission is an international collaboration between \href{https://www.esa.int/Science_Exploration/Space_Science/LISA}{ESA}, its member states, and \href{https://lisa.nasa.gov/}{NASA}, for the detection of gravitational waves from space \cite{colpi2024lisadefinitionstudyreport}. 
It was adopted in January 2024 and is scheduled for launch in the mid-2030's. It will be a constellation of three identical spacecraft forming a near-equilateral triangle in an heliocentric orbit, transferring laser beams over $2.5 \cdot 10^6$ km long arms. Laser interferometry is used to track separations between test masses, thus measuring spacetime strain variations as a function of time. 
LISA Science Objectives tackle many open questions in astrophysics, fundamental physics and cosmology, including ESA's Cosmic Vision questions \cite{CosmicVision} ``What are the fundamental laws of the universe?'' and ``How did the universe originate and of what is it made?''.
%LISA will use precision laser interferometry to compare separations between test masses that are protected by the spacecraft from non-gravitational disturbances. Thus, it will coherently measure spacetime strain variations as a function of time, including frequency, phase, and polarisation, forming a continuous, all-sky observatory sensitive to all sources simultaneously. LISA’s science objectives cover a wide range of outstanding questions in astrophysics, fundamental physics and cosmology, including ESA’s Cosmic Vision questions \cite{CosmicVision} such as ``What are the
%fundamental laws of the universe?'' and ``How did the universe originate and of what is it made?''.
%The LISA science output will have outstanding synergy with particle physics; 
In this contribution, based on the LISA Red Book \cite{colpi2024lisadefinitionstudyreport}, we present a summary of the LISA Science Objectives relevant for the European Strategy for Particle Physics.
\end{abstract}

\vspace*{\fill}
${^*}$Corresponding author: \href{mailto:chiara.caprini@cern.ch}{chiara.caprini@cern.ch}

\newpage

\clearpage
\setcounter{page}{1}

\section{Mission overview}

% ========================================================================

\begin{figure}[ht]
    \centering
    \vspace{-0.2cm}
    \includegraphics[width=0.8\textwidth]{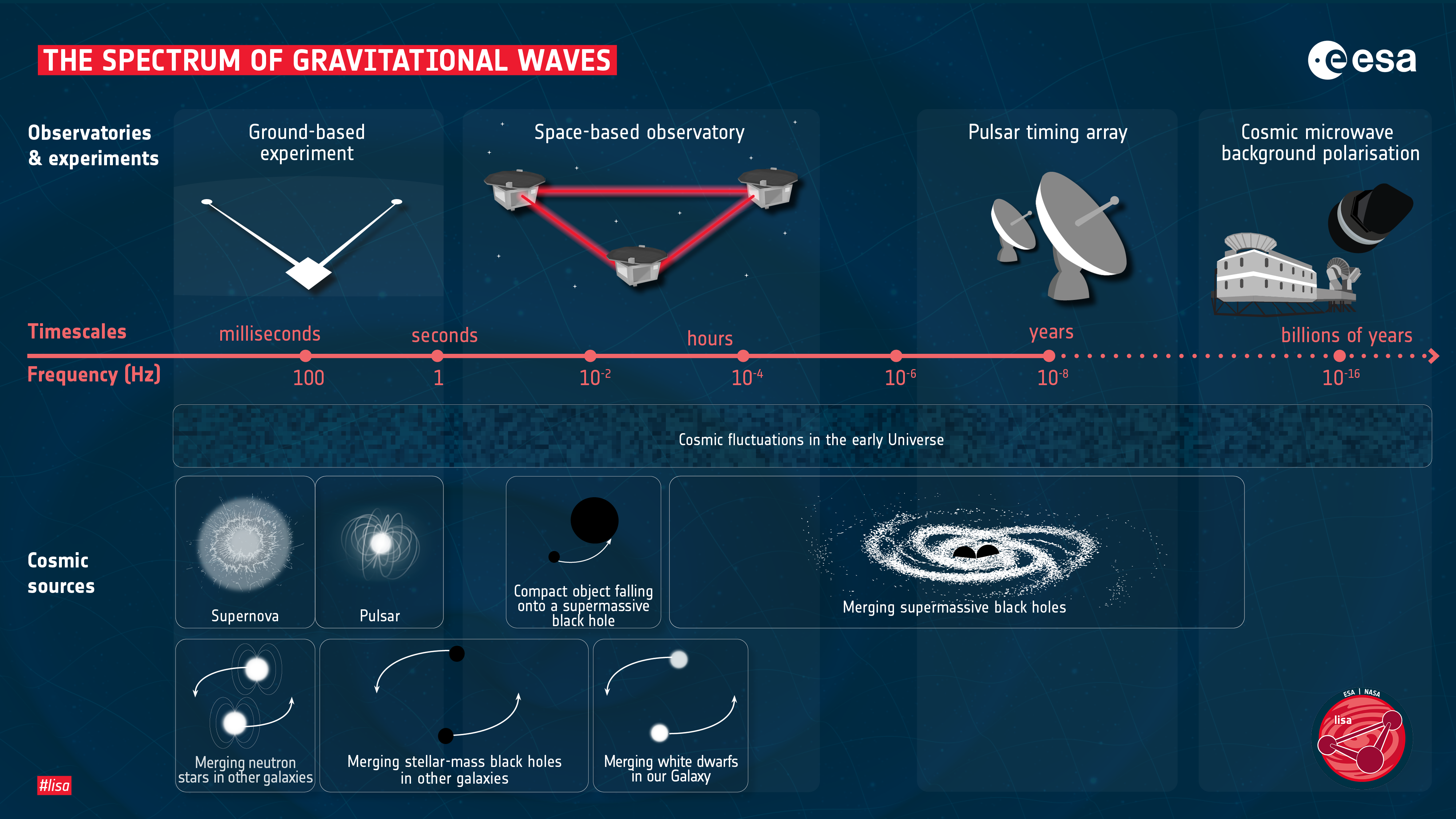}\\
    \caption{\footnotesize Taken from \cite{colpi2024lisadefinitionstudyreport}. LISA targets the milliHz band of gravitational waves, lying between the nanoHz
      regime probed by Pulsar Timing Arrays and the decaHz regime accessible to ground-based detectors. Several types of sources produce gravitational waves in this band, which also corresponds to the TeV energy scale in the early universe: LISA can thus access gravitational wave signals linked to beyond Standard Model processes.}
    \label{fig:gw_spec}
\end{figure}

{Gravitational Waves (GWs)} are a unique probe, with immense discovery potential.
They were %only recently 
first directly detected in %September 
2015 by the LIGO-Virgo collaboration \cite{LIGOScientific:2016aoc}. This ground-breaking initial detection gave way to many follow-on scientific discoveries, %e.g., 
such as the discovery of a %previously unknown 
new class of black holes (BHs)~\cite{LIGOScientific:2021qlt}; the first GW-electromagnetic (EM) multimessenger observation, 
confirming the link between neutron star mergers and short gamma ray bursts~\cite{2017ApJ...848L..12A}; 
and tight constraints on beyond General Relativity (GR) theories~\cite{LIGOScientific:2020tif}.
In 2023, Pulsar Timing Arrays (PTAs) Collaborations worldwide announced the first evidence for a low frequency (nanoHz) GW background \cite{EPTA:2023fyk,NANOGrav:2023gor,Reardon:2023gzh, Xu:2023wog, InternationalPulsarTimingArray:2023mzf}. 
This measurement potentially provides the first proof of super-massive black hole binaries existing and merging in the universe, overcoming the last-parsec problem. It might also constitute the first ever detection of a fossil GW radiation from the primordial universe. 

\subsection{Gravitational wave sources in the LISA band}
\label{sec:sources}

Different GW sources emit at different frequencies and, as with EM radiation, different science opportunities reside in different bands of the GW
spectrum (see \cref{fig:gw_spec}).
The mHz band, residing between the LIGO-Virgo-KAGRA (LVK) and PTA {bands}, is expected to host a diverse population of sources, making it ideal for GW astronomy. \emph{\color{red}{The LISA space mission aims to unlock this unexplored band.}}

\begin{figure}
    \centering
    \vspace{-0.5cm}
    \includegraphics[width=0.8\textwidth]{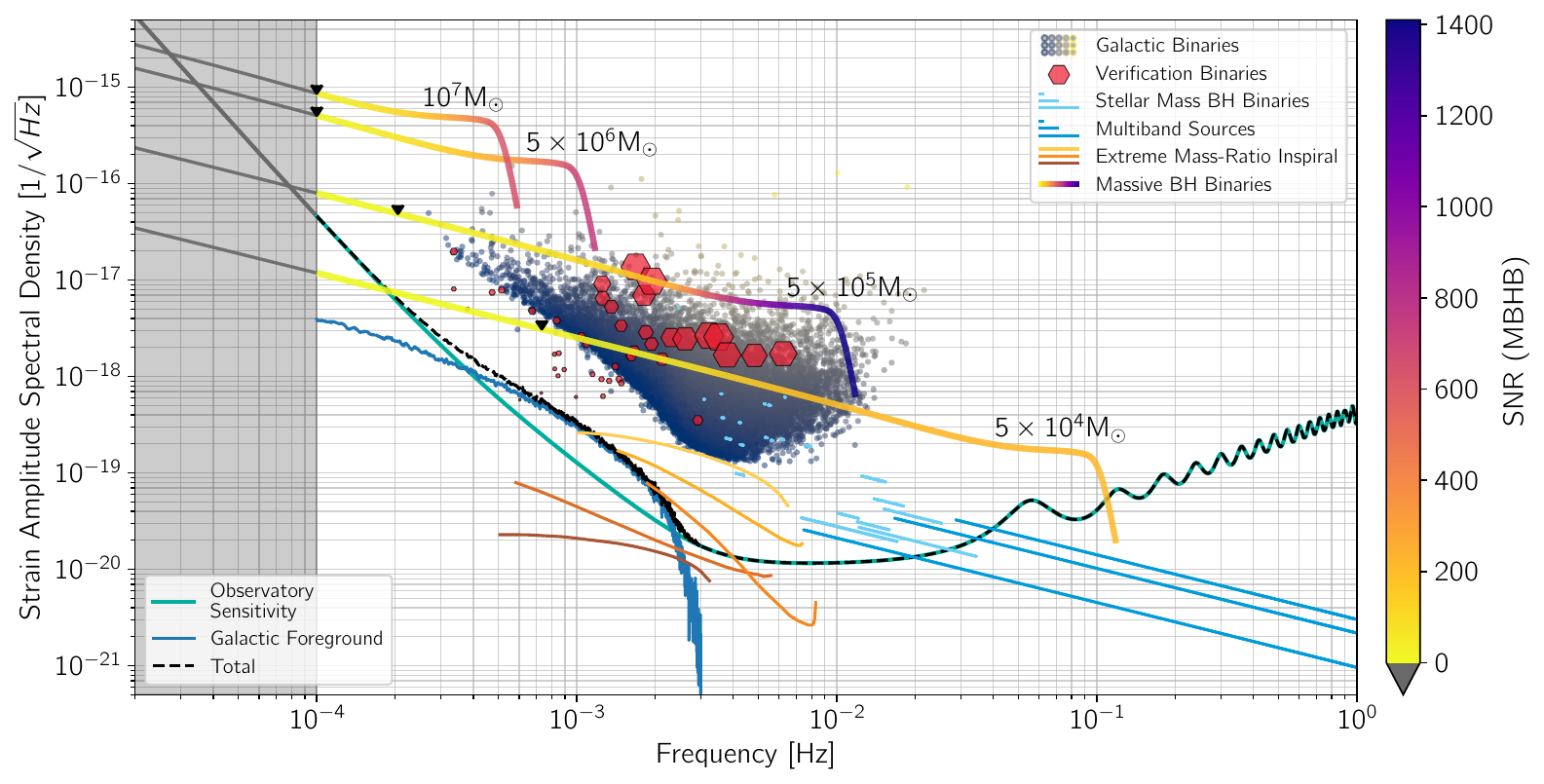}\\
    \caption{\footnotesize Taken from \cite{colpi2024lisadefinitionstudyreport}. Illustration of the primary LISA source classes in the GW
      frequency-amplitude plane. Included are merging massive black hole binaries; an extreme mass ratio inspiral at moderate
      redshift; stellar mass black hole binaries at low redshift, including potential multiband sources; galactic binaries, including
      verification binaries in the Milky Way. Solid teal, solid blue and dashed black lines denote sensitivity
    limits from instrumental noise alone, the unresolved GW foreground, and their sum, respectively.
    The grey shaded area is the extrapolation of LISA's instrumental noise below {0.1} {mHz}. 
    }
    \label{fig:gw_sources}
\end{figure}

\cref{fig:gw_sources}, taken from \cite{colpi2024lisadefinitionstudyreport},
illustrates LISA sources in the GW frequency vs.~amplitude plane, along with the nominal mission lifetime instrumental noise. 
Working upward, the teal line shows LISA noise requirements as equivalent GW strain. 
The dashed black line adds the unresolved GW foreground from persistent Milky Way sources (blue line), which limits LISA sensitivity to resolved sources at ${0.3} \,{\rm mHz}\lesssim f \lesssim {3} \,{\rm mHz}$, but probes Galactic sources.

%provides an illustration of a selection of LISA sources in the {GW} frequency vs.\
%amplitude plane, as well as a representation of the instrumental noise assuming nominal mission lifetime. 
%Working upwards from the bottom of \cref{fig:gw_sources}, the teal line shows LISA instrument noise %requirements
%expressed as equivalent {GW} strain. The dashed black line adds
%the unresolved {GW} foreground (blue line) from persistent sources in the Milky Way, which are not individually
%identifiable. %by LISA. 
%This foreground limits LISA sensitivity to resolved sources in the frequency band
%${0.3} \,{\rm mHz}\lesssim f \lesssim {3} \,{\rm mHz}$, but is a probe of Galactic binaries.
%sources.

\emph{\color{red}{Extreme Mass Ratio Inspirals (EMRIs)}} result from the orbit of a stellar mass black hole (sBH) around a much more massive BH and can be detectable for months to years. 
They produce complex waveforms with multiple harmonics of the fundamental GW frequency, detectable by LISA in its sensitivity bucket, at $\qty{1}{mHz}\lesssim f \lesssim \qty{10}{mHz}$. 
%EMRIs are long-lasting sources with timescales of several years. 
The example EMRI signals in \cref{fig:gw_sources} (orange lines) show the fundamental and first four harmonics from the inspiral of a $\qty{35}{\Msolar}$ {BH} into a $\qty{2e5}{\Msolar}$ {BH} at redshift $z=2$. %\dani{Which of those lines is that one?} \CC{all of them} 
%\DMS{Are you going to indicate the color of their lines like above or was that just for the noise? } CC: COLOR OF LINES ADDED
EMRIs can probe the stellar population in galaxies, measure the mass and spin of the central massive BH, and map the spacetime metric surrounding it, testing the no-hair theorem and theories beyond Einstein.

%Near the bottom of LISA's sensitivity {bucket}, the Extreme Mass Raio Inspirals (EMRIs) radiate between
%$\qty{1}{mHz}\lesssim f \lesssim \qty{10}{mHz}$. {EMRIs}, which result from the orbit of a 
%stellar mass black hole (sBH) around a companion, much more massive, BH (MBH), produce complex waveforms with multiple
%harmonics of the fundamental {GW} frequency, simultaneously detectable by LISA. {EMRIs} are long-lasting
%sources with timescales of several years. The example {EMRI}
%signals in \cref{fig:gw_sources} shows the fundamental and first four harmonics from the inspiral of 
%a $\qty{35}{\Msolar}$ {BH} into a $\qty{2e5}{\Msolar}$ {BH} at redshift $z=2$. \dani{Which of those lines is that one?} \CC{all of them}
%EMRIs can be used to probe the stellar population in the central regions of galaxies and measure the mass and spin of the central MBH. 
%They are also outstanding sources to map the detailed structure of the spacetime metric of the central {MBH}, allowing
%tests of the no-hair theorem, the existence of additional fields and other alternative Gravity theories.

\emph{\color{red}{Binary stellar-mass black holes (sBHBs)}}, the type of BHB that is regularly detected by LVK, can also be detected by LISA in the nearby universe during their early evolution. 
These are represented by the cyan lines at frequencies higher than those of EMRIs.
Some sBHBs (darker lines) reach the merger phase in a time frame close to LISA observations, evolving into the sensitivity band of ground-based GW instruments. This enables \emph{multiband} GW astronomy, namely, detecting the same system in multiple frequency bands with different GW instruments. Those that merge further out (hence undetectable by ground detectors) are represented by the lighter lines.

\emph{\color{red}{Galactic binaries}}, the most plentiful LISA sources, are binary systems in the Milky Way composed of two compact objects with periods of hours to minutes. LISA is expected to detect over ten thousand individual sources during its nominal science phase. 
These sources, mostly stable, are represented by points in the figure and provide insight into the final stages of stellar binary evolution, the tides and mass transfer, and the Galactic structure. 
Some systems' slow frequency evolution will be measured in the time frame of LISA observations. %, allowing direct estimation of their luminosity distance. %using the standard siren technique. 
The red hexagons indicate the verification binaries, a subset of Galactic binaries already identified through their EM emission: these guaranteed multimessenger sources offer additional science opportunities and can be used to verify LISA on-orbit performance. The hexagons size indicates the system {signal-to-noise ratio (SNR)}.

\emph{\color{red}{Massive black hole binaries (MBHBs)}}, the loudest individual signals in LISA, live in the upper half of \cref{fig:gw_sources}. 
Four representative systems are shown with
different masses and at $z=3$ (gradient-coloured lines from yellow to violet, representing SNR).  
LISA will be sensitive to the late inspiral, merger, and ringdown of MBHBs with masses ranging from $\qty{e3}{\Msolar}\lesssim M \lesssim \qty{e7}{\Msolar}$ at redshifts $z\lesssim 15$, 
directly probing the origin of massive BH and their growth
across the universe. 
Some of these sources will have
SNR as large as 1000, providing exquisite tests of GR and of the nature of event horizons. 
Multi-messenger observations will provide additional information about matter surrounding the mergers, can be used to test the propagation of {GWs}, and to construct the Hubble diagram up to high redshift. 

The superposition of low-SNR, overlapping GW signals from various sources can appear as stochastic signals in LISA. 
Galactic binaries (blue line) are the only ones shown in \cref{fig:gw_sources}, but extra-galactic sources like EMRIs, sBHBs, White Dwarf binaries, and even MBHBs (for some formation models), can also generate \emph{\color{red}{stochastic astrophysical foregrounds}}. 
Furthermore, potential GW sources in the early universe create \emph{\color{red}{stochastic GW backgrounds (SGWBs)}}, since
they are homogeneous and isotropic over the sky, and/or correlated on scales much smaller than the
detector resolution. 
\emph{\color{red}{If detected, SGWBs %\gf{in the mHz frequency window} 
would provide direct access to the first microsecond of the universe's lifetime, carrying imprints of fundamental physics at high energy scales and potentially revealing new insights into high-energy particle interactions, %\gf{around and above the TeV energy scale}, 
in a complementary way to present and future particle colliders}}.
%\dani{There aren't any citations really for these curves, don't know if it's an idea to cite some paper where all the phtysics in figure 2 explained}
%\CC{citations are infinite: I have just cited the red book at the beginning}

\subsection{The LISA mission concept}

%\begin{figure}
%    \centering
%    \vspace{-0.5cm}
%\includegraphics[width=\textwidth]%{mission_summary.png}\\
%    \label{fig:mission}
%\end{figure}

LISA aims to achieve its science goals with a 4.5-year observation period using orbits that maintain the triangular constellation of three spacecraft for about a decade without active station keeping. 
ESA leads the mission, providing the spacecraft, launch, operations, and data handling. 
NASA contributes the ultra-stable lasers, telescopes, and UV light source for test mass discharge. 
ESA’s member states provide the rest of the instrumental elements. 

A fully optical chain of beam-splitting, reflection, and recombination, as seen in ground detectors, is unfeasible over the 2.5 million km baseline. 
Instead, each spacecraft emits a one-way link, with its own laser source, to the other two spacecraft; the interferometric measurement is made between outgoing and incoming light from a distant spacecraft. 
The recorded phases are combined with \emph{Time Delay Interferometry} (TDI) to remove noise \cite{Tinto:2020fcc,2005PhRvD..71b2001V,2022arXiv221205351B, 2010CQGra..27h4012M, 2012PhRvD..86l2006M,
  2015ExA....39..281G, 2010PhRvL.104u1103D}. 

The observatory generates continuous time series of TDI observables sensitive to GW across the frequency band and from all sky directions. 
These measurements are transmitted to Earth, where the ESA Science Operation Centre transforms the data from Level-0 (L0, raw data), through L0.5 (that are synchronised, in physical units and usable format), 
to L1, removing major noise sources. 
Signals from tens of thousands distinct sources will be superimposed in L1 data. These will be extracted and characterised 
at the ESA Distributed Data Processing Centre and the NASA Science Ground Segment by 
simultaneously using a global fit procedure \cite{Littenberg:2023xpl,Rosati:2024lcs,Katz:2024oqg,Deng:2025wgk}, resulting in a collection of posterior parameter distributions for identified sources (L2 data). 
The residuals from the subtraction procedure will be searched for unmodelled astrophysical or cosmological signals, and used to characterise the observatory’s performance.
In a final step, the source parameter distributions from both analyses will be combined into a single catalogue of GW source candidates (L3 data) with detection confidence and estimated astrophysical parameters.
In a parallel ``low latency'' analysis process, transient events will be detected and alerts issued for relevant sources as soon as they appear, allowing EM observations of the event counterpart.

LISA will release L0.5 to L3 data regularly to the scientific community. 
LISA catalogues will provide direct access to masses, spins, distances, and sky positions for systems of compact objects across a wide mass and distance range. 
This information will help {pin down source models and enable (multimessenger) astrophysics, fundamental physics, and cosmology.}

\section{Science overview}

The science of LISA is broad and diverse, and has been described in a series of white papers \cite{2022LRR....25....4A,LISACosmologyWorkingGroup:2022jok,LRR23,2022GReGr..54....3A,2019BAAS...51c..34L,2019BAAS...51c..67C,2019BAAS...51c.432C,2019BAAS...51c.109C,2019BAAS...51c.123B,2019BAAS...51c..73N,2019BAAS...51c..42B,2019BAAS...51c..76C,2019BAAS...51c.455M,2019BAAS...51c..32B,2019BAAS...51c.175B}.  
The mission Science Objectives (SOs), listed in \cref{tab:mission}, are in relation with the classes of sources described in \cref{sec:sources}. 
In the following, we review the SOs that, we believe, are more connected to the focus of the European Strategy for Particle Physics.

\subsection{SO5: Explore the fundamental nature of gravity and Black Holes}
Several sources for LISA (MBHBs, EMRIs, IMRIs) have the potential for detectable GW signals with thousands of cycles and, in some cases, extremely high SNRs. This enables highly precise measurements of several parameters, which can be used to \emph{\color{red}{constrain both extensions of GR and the Standard Model (SM), as well as probe dark matter and dark energy models}} \cite{2022LRR....25....4A}. In addition, tracking sBHBs through the LISA frequency band and later in the ground-based detector's band enables one to capture the evolution in radiation and search for non-GR or beyond SM effects. In particular:
\begin{itemize}
    \item The post-merger ringdown of a MBHB can determine whether the objects are indeed black holes as predicted by GR. %\RB{I propose to say instead ``The post-merger ringdown of a MBHB can determine whether the objects are indeed the black holes predicted by GR.'', since you can also test beyond GR black holes besides non-BHs}.
    \item EMRIs can be used to probe the multipolar structure of massive BHs, testing the no-hair conjecture and the BH nature of supermassive compact objects with very high accuracy. %\RB{I would change this sentence to ``EMRIs can be used to probe the multipolar structure of MBHs, testing the no-hair conjecture and the BH nature of supermassive compact objects with very high accuracy.'', since I think that's more accurate and generic. The part regarding probing new light fields with EMRIs is not so much related to the multipolar structure but rather to additional emission channels that is discussed in the next bullet point (in some cases you might have both deviations in the multipolar structure and have additional channels of emission, but in any case I would keep the sentence more generic by saying something like I suggested)}
    \item Using MBHBs, EMRIs and multiband sBHBs, one can search for beyond vacuum GR emissions or polarisations, which could determine the presence of e.g. scalar hair.
    \item MBHBs, EMRIs and multiband sBHBs can be used to test the propagation properties of GWs, in particular whether they respect Lorentz symmetry and parity invariance.
\end{itemize}

%
%
%
%%%%%%%%%%%%%%%%%%%%%%%%%%%%%%
%%%%%%%%%%%%%%%%%%%%%%%%%%%%%%
%
\begin{minipage}[t]{0.95\textwidth}
\begin{center}
%\vspace{2.0em}
\fontsize{10pt}{11}{\bf LISA MISSION SUMMARY}
\end{center}
\footnotesize
\begin{tabular}{|p{\textwidth}|}
\hline
\rowcolor{DeepSpace}
\multicolumn{1}{|c|}{\textcolor{White}{\textbf{Science Objectives (SO)}}}\\
\hline
\begin{itemize}
    \item[{\bf SO1:}] Study the formation and evolution of \textbf{compact binary stars} and the structure of the Milky Way Galaxy
    \item[{\bf SO2:}] Trace the origins, growth and merger histories of \textbf{massive black holes} across cosmic epochs
    \item[{\bf SO3:}] Probe the properties and immediate environments of Black Holes in the local universe using \textbf{{Extreme}} and \textbf{{Intermediate}} mass ratio inspirals
    \item[{\bf SO4:}] Understand the astrophysics of \textbf{{stellar black holes}}
\end{itemize}
%\vspace{1.0em}
%
\begin{minipage}{0.75\textwidth}
\begin{itemize}
    \item[{\bf SO5:}] Explore the \textbf{fundamental nature of gravity} and Black Holes
    \item[{\bf SO6:}] Probe the rate of \textbf{expansion of the universe} with standard sirens
    \item[{\bf SO7:}] Understand \textbf{stochastic {GW} backgrounds} and their implications for the early universe and TeV-scale particle physics
    \item[{\bf SO8:}] Search for {GW} bursts and \textbf{unforeseen sources}
\end{itemize}
\vspace{1.0em}
 \end{minipage}
\hfill
\begin{minipage}{0.228\textwidth}
 \vspace{-3em}
\centering
\includegraphics[width=0.75\textwidth]{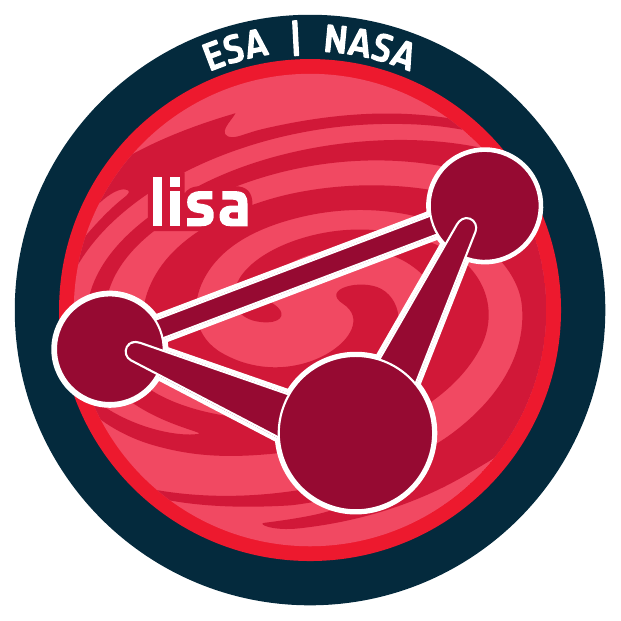}
 \end{minipage}\\
\hline
\end{tabular}
\vspace{-1.5em}
\begin{center}
    \footnotesize
    \begin{tabular}{|m{\textwidth}|}
        \hline
        \rowcolor{DeepSpace}\multicolumn{1}{|c|}{\textcolor{White}{\textbf{Measurement}}}\\
        \hline
        \vspace{0.1em}
        {GW} in the Frequency Band of \qty{0.1}{mHz} - \qty{1.0}{Hz} with a {GW} Strain
        Spectral Density: \numrange{e-21}{e-23} \\ [0.3em]
        \hline
    \end{tabular}
\end{center}
\vspace{-1.5em}
\begin{center}
\footnotesize
\begin{tabular}{|m{0.28\textwidth}|m{0.695\textwidth}|}
    \hline
    \rowcolor{DeepSpace}\multicolumn{2}{|c|}{\textcolor{White}{\bf{Payload}}}\\
    \hline
    Lasers & 2 per spacecraft
    $\bullet$ \qty{2}{W} output power
    $\bullet$ wavelength \qty{1064}{nm}
    $\bullet$ frequency stability \qty{300}{\hertz/\rtHz}\\
    \hline
    Optical Bench & 2 per spacecraft
    $\bullet$ double-sided use
    $\bullet$ high thermal stability ({Zerodur})\\
    \hline
    Interferometry & heterodyne interferometry
    $\bullet$ \qty{15}{\pico\meter/\rtHz} precision
    $\bullet$ Inter-spacecraft ranging to \qty{\sim 1}{m}\\
    \hline
    Telescope & 2 per spacecraft
    $\bullet$ \qty{30}{cm} off-axis telescope
    $\bullet$ high thermal stability\\
    \hline
    Gravitational Reference System & 2 per spacecraft
    $\bullet$ acceleration noise \qty{<3}{\femto\meter/(\second\squared\rtHz)}
    $\bullet$ \qty{46}{mm} cubic AuPt test mass
    $\bullet$ Faraday cage housing
    $\bullet$ electrostatic actuation in 5 {degrees of freedom}  \\
    \hline
\end{tabular}
\end{center}
\vspace{-2em}
\begin{center}
\footnotesize
\begin{tabular}{|m{0.11\textwidth}|m{0.865\textwidth}|}
    \hline
    \rowcolor{DeepSpace}\multicolumn{2}{|c|}{\textcolor{White}{\bf{Mission}}}\\
    \hline
    Duration & \qty{4.5}{years} science orbit
    $\bullet$ \qty{>82}{\%} duty cycle
    $\bullet$ \qty{\sim 6.25}{years} including transfer and commissioning\\
    \hline
    Constellation  & Three drag-free satellites forming an equilateral triangle
    $\bullet$ \qty{2.5e6}{km} separation
    $\bullet$ trailing/leading Earth by \ang{\sim20}
    $\bullet$ inclined by \ang{60} with respect to the
    ecliptic \\
    \hline
    Orbits  & Heliocentric orbits
    $\bullet$ semimajor axis \qty{\sim1}{AU}
    $\bullet$ eccentricity $e \approx 0.0096$
    $\bullet$ inclination $i \approx \ang{0.96}$\\
\hline
\end{tabular}
\end{center}
\vspace{-2em}
\begin{center}
\footnotesize
\begin{tabular}{|m{0.1\textwidth}|m{0.075\textwidth}|m{0.775\textwidth}|}
    \hline
    \rowcolor{DeepSpace}\multicolumn{3}{|c|}{\textcolor{White}{\bf{Data Analysis}}}\\
    \hline
    Noise \newline Reductions & \multicolumn{2}{m{0.865\textwidth}|}{Laser noise suppression
    with time-delay interferometry
        $\bullet$ Ranging processing and delay estimation
        $\bullet$ spacecraft jitter suppression and reduction to 3 lasers
        $\bullet$ Tilt-to-length effect correction
        $\bullet$ Clock noise suppression
        $\bullet$ Clock synchronisation}\\
    \hline
     & L0  & Raw data, de-multiplexed, time-ordered, corruption removed \\
    & L0.5 & Primary science telemetry, decommutated, time-stamped, unit-level calibrations applied \\
     Data \newline Levels & L1 & Time Delay Interferometry variables ({GW} strain) \\
    & L2 & Output from a global fit pipeline, statistical evidence for candidate sources \\
    & L3 & Catalogue of {GW} sources (detection confidence, estimated astrophysical parameters) \\
    \hline
\end{tabular}
\end{center}
\vspace{-1em}
\footnotesize 
\captionof{table}{\footnotesize Taken from \cite{colpi2024lisadefinitionstudyreport}. Science objectives, characteristics and data products of the LISA mission.}
\label{tab:mission}
\end{minipage}
\vspace*{1.0em}
%
%\arrayrulecolor{Black}
%
%%%%%%%%%%%%%%%%%%%%%%%%%%%%%%%%%%%%%%%%%%%%%%%%%%%%
%%%%%%%%%%%%%%%%%%%%%%%%%%%%%%%%%%%%%%%%%%%%%%%%%%%%
%
%
%

The post-merger ringdown of a MBHB sees the decline/ringdown of several modes which, according to GR, are completely described by the remnant's mass and dimensionless spin \cite{Echeverria:1989hg}. Detection of one mode can therefore be used to infer these properties, while more modes will increase the accuracy of such a measurement \cite{Berti:2005ys}. Exotic systems involving boson stars, fuzzball/microstate scenarios, quantum-gravity motivated extensions of GR, presence of dark matter etc.~would appear or lead to horizonless objects, which in turn gives rise to a severely 
modified oscillation mode spectrum \cite{Cardoso:2016oxy,Berti:2018vdi, Dreyer:2003bv}. 
To confirm the Kerr nature of the remnant (and rule out such exotic systems), detection of at least three ringdown modes are required with SNR of 8 or higher, which is very much within LISA's capabilities. 
%for MBHBs.

EMRIs %, systems expected to lie at the centre of galaxies, 
can have $10^5$ orbits within the LISA band, and in some cases with a SNR $>50$. Many orbits translates to a pristine relative accuracy (e.g. $10^5 \text{ orbits} \Rightarrow 10^{-5}$ relative accuracy); this in turn allows one to probe the quadrupole moment of Kerr to $10^{-4}$ \cite{2017PhRvD..95j3012B}. Should either body deviate from a Kerr BH significantly via hair (quantities other than mass and spin leading to e.g. scalar or vector radiation) \cite{Doneva:2022ewd,Eichhorn:2023iab}, the presence of a light boson cloud \cite{Arvanitaki:2014wva,Herdeiro:2014goa}, or simply not be a BH \cite{Vaglio:2022flq,Bianchi:2020bxa}, this could be detectable; meanwhile a non-detection enables constraints.

Pure GR waveforms have two polarisations, however should there be other emission channels, leading to scalar or vector radiation, this also leads to scalar and vector polarisations, to which LISA will be sensitive for MBHBs and EMRIs \cite{Tinto:2010hz}. In addition, any EMRI detection \cite{2022NatAs...6..464M,Barsanti:2022vvl} or multiband detection of sBHBs \cite{Moore:2019pke} will see the binaries evolve differently should there be these `extra' forms of radiation or if the binaries evolved in a non-vacuum environment, e.g.~accretion disks or dark matter environments~\cite{2021PhRvL.126j1105T,Eda:2013gg,Duque:2023seg}. If sBHBs detected in LISA are picked up at a later time %in the 
by ground-based detectors, the evolution of their inspirals can inform whether other forms of radiation have occurred \cite{Barausse:2016eii, Carson:2019rda}. Such multiband observations can also break degeneracies between GR and non-GR parameters.

Any deviation from the expected propagation properties of GWs, albeit speed, phase or amplitude, can have consequences in fundamental physics \cite{Chamberlain:2017fjl}. Models of dark energy can predict frequency-dependent deviations in the GW speed \cite{deRham:2018red}, making it important to extend the existing ground-based detector constraints \cite{LIGOScientific:2017ync} to lower LISA frequencies. Changes in the GW speed can be constrained to $10^{-15}$ with e.g.~a multiband, GW150914-like sBHB \cite{Baker:2022eiz}. 
The GW phase can be used to tighten constraints on the graviton mass by 2-3 orders of magnitude than the current LVK limits \cite{Perkins:2020tra}. Lorentz-violating theories like Einstein-Aether can also impact the phase and would be detectable in EMRIs \cite{Chamberlain:2017fjl}. Finally, a dampening in the amplitude over large distances in MBHB signals would be seen in several dark energy scenarios \cite{LISACosmologyWorkingGroup:2019mwx}.

\subsection{SO6: Probe the rate of expansion of the universe with standard sirens}

The GW emission from a binary provides a direct measure of its luminosity distance, unlike distant EM sources like type-Ia supernovae for which calibration with local rulers is necessary.
With an associated redshift measurement, GW sources can therefore be used to construct the Hubble diagram, and probe the background expansion of the universe \cite{Schutz86}.  
However, GW measurements alone do not provide the source redshift: to determine it, one needs to rely on measurements of different type. 
This has so far been achieved with LVK data both by the observation, in several EM bands, of the EM counterpart of a NS merger event \cite{2017ApJ...848L..12A,LIGOScientific:2017adf}, and by cross-correlating GW events with galaxy catalogues \cite{LIGOScientific:2018gmd,DES:2019ccw,LIGOScientific:2021aug,LIGOScientific:2019zcs,Gray:2023wgj,Mastrogiovanni:2023emh}. \textit{Bright sirens} and \textit{dark sirens} is the popular nomenclature to distinguish these two approaches.

Concerning LISA sources, detectable EM counterparts are expected for MBHB mergers at the centre of galaxies, if surrounded by gas: 
the redshift of the GW source is then provided by the identification of the host galaxy \cite{Tamanini16,Lops2022,Mangiagli2022}.
Other distant sources like EMRIs and sBHBs are not expected to have detectable counterparts, but EMRIs should be numerous enough to be used as dark sirens \cite{Laghi21}.
\emph{\color{red}{LISA will measure the universe’s expansion rate across cosmic history: using MBHB as {bright sirens}, and EMRIs as {dark sirens}, LISA can probe the remarkably wide redshift range $0.01 \lesssim z \lesssim 7$.}}

\vspace{0.5em}
$\bullet$~\textbf{Bright sirens, MBHBs}. 
The main advantage of MBHBs as bright sirens is that they are numerous at high redshift: multimessenger observations of {MBHB} mergers in the \qtyrange{e5}{e6}{M_{\odot}} mass range are possible up to $z\lesssim 7$.
In 4 years of LISA observation, Ref.~\cite{Mangiagli2022} predicts between $\sim7$ and $\sim20$ {MBHB} bright sirens, based on a detection strategy involving future {EM} facilities across different bands: \href{https://www.vro.org/}{VRO}, \href{https://www.skao.int/en/explore/telescopes}{SKA}, \href{https://elt.eso.org/}{ELT}, and \href{https://www.cosmos.esa.int/web/athena}{Athena}. 
The best constraints on the Hubble rate $H(z)$ are obtained in the range $1.5\lesssim z\lesssim 3$, where the GW events cluster.
LISA will be able to constrain $H(z=2)$
%\NT{Better to say ``the Hubble rate at $z=2$'' to be clearer} CC: it is defined a couple of lines above
with an accuracy better than \qty{10}{\%}: a unique opportunity to test the evolution of the universe at high redshift, still poorly traced by {EM} observations, where deviations from $\Lambda$CDM~\cite{Speri21,Caprini16,2019JCAP...07..024B} may thus hide.

\vspace{0.5em}
$\bullet$~\textbf{Dark sirens, EMRIs}: 
EMRIs detectable by LISA are distributed at lower redshift than MBHBs: as dark sirens, they are therefore complementary to MBHBs, and offer the opportunity to measure the Hubble constant $H_0$, and the equation of state of dark energy $w_0$.
The measurement precision depends on the number of detected {EMRIs}, still highly uncertain: between a few to several thousands per year out to $z\sim 3$, depending on population models  \cite{2017PhRvD..95j3012B}.
Based on a fiducial EMRI population, and assuming galaxy catalogues complete up to $z=1$, Ref.~\cite{Laghi21} showed that $H_0$ can be determined at few percent, while the uncertainty on $\Omega_M$ is large; fixing $H_0$ and $\Omega_M$, $w_0$ can be determined with an error around 10\%, and deviations from GR can be constrained at few percent \cite{Liu:2023onj}.

\vspace{0.5em}
The joint-inference on $H_0$ and $\Omega_M$ obtained by combining EMRIs and MBHBs is expected to improve the constraints to the sub-percent level on $H_0$, and a few percent on $\Omega_M$ \cite{Tamanini:2016uin}.
%\nt{I moved this last sentence outside the EMRIs paragraph and added a ref.}

% ************************************************************************
% ========================================================================

\subsection{SO7: Understand stochastic GW backgrounds and their implications for the early universe and TeV-scale particle physics}

{GW} sources in the early universe form a fossil radiation carrying information on the first instants of the universe, and thereby on high energy physics.
Remarkably, {the LISA frequency band encompasses the TeV energy scale}, the frontier of our knowledge of the theory of fundamental interactions. 
Indeed, the characteristic frequency of a SGWB of primordial origin can be related to the Hubble scale $H_*$ at the moment of GW generation, $f_*\geq H_*$, so that the frequency today is 
\begin{align}
    f &= f_* \frac{a_*}{a_0} \simeq  {1.65}\times 10^{-4}\,{\rm Hz}\,\left(\frac{g_*}{100}\right)^{\frac{1}{6}} \frac{f_*}{H_*}\,\frac{T_*}{{1}\,{\rm TeV}} \;, 
    \label{redsh_H}
    \end{align}
%    \GN{Autistic comment: typically people put a small space between the equation and the final comma, like \;}
where $a_{*},\,a_0$ denote the scale factor,  $g_*$ the number of relativistic degrees of freedom, and $T_*$ the universe temperature.
The LISA frequency window $[10^{-4},0.1]$ Hz {hence} provides access to sources operating when the universe had a temperature around and beyond the TeV scale. 
\emph{\color{red}{LISA has therefore discovery potential in complementarity with the LHC and future colliders, which are scheduled on a longer time-scale than LISA}}. 

\begin{wrapfigure}{r}{0.45\textwidth}
\includegraphics[width=0.45\textwidth]{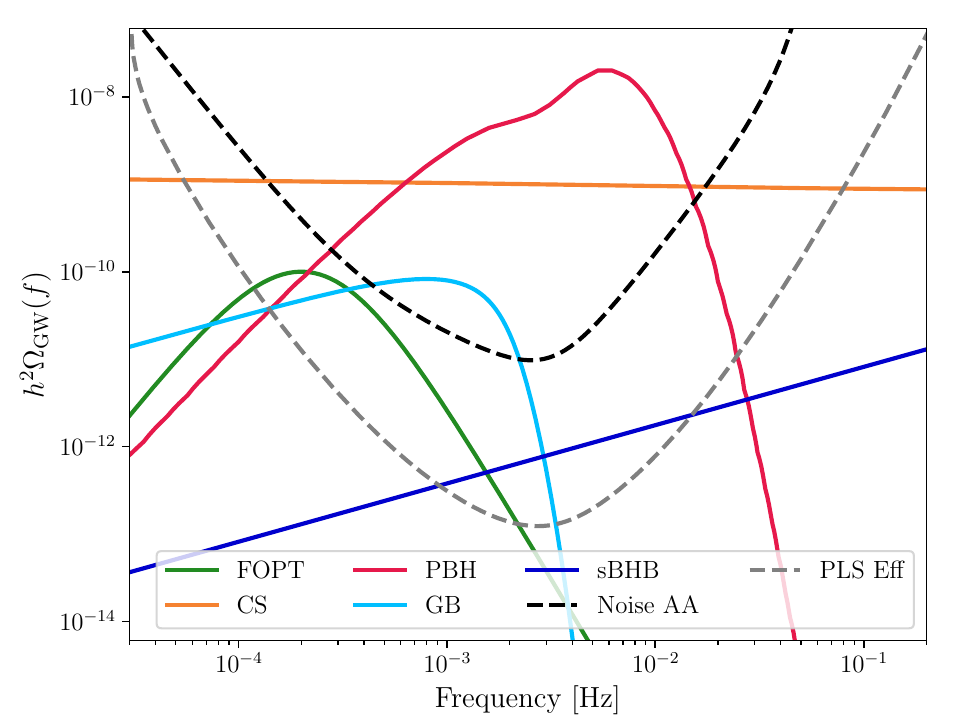}
    \caption{\footnotesize Taken from \cite{colpi2024lisadefinitionstudyreport}. Examples of {SGWBs}
    in the LISA band, together with the instrument sensitivity in the $A$-channel ({black, dashed}) and the
    effective Power Law Sensitivity \cite{Thrane:2013oya} ({grey, dashed}). The {SGWBs} are: in {red}, the {SGWB} associated to the formation of {primordial black holes} in a mass range for which they could constitute the totality of the dark matter \cite{Bartolo:2018rku};
    in {orange}, the {SGWB} from {cosmic strings} with tension providing a signal that would account for the {SGWB} detection by {PTAs} \cite{2023arXiv230616227A};  in {green}, the
    {SGWB} from a primordial first order phase transition at the {EW} scale, in the context of the SM singlet extension, testable at particle colliders. The astrophysical {SGWB} from unresolved {sBHBs}, taken from \cite{SOBBHtoappear} assuming GWTC-3 population constraints \cite{2021arXiv211103634T} is shown in {dark blue}. The Galactic foreground is shown in {light blue}, taken from
    \cite{Karnesis:2021tsh}, averaged over time.
    }\label{fig:allSGWB}
\end{wrapfigure} 

Furthermore, LISA is sensitive to many possible {GW} sources operating at a variety of energy scales in the early universe (see \cref{fig:allSGWB}), with the potential of clarifying, in complementarity with cosmological observations such as the Cosmic Microwave Background (CMB) and Large Scale Structure, some outstanding phenomena in cosmology: 
the nature of dark matter, the origin of the matter-antimatter asymmetry of the universe, the scale of Grand Unification, inflation
%} 
\cite{Caprini:2015zlo,Caprini:2024hue,coswg_PE_phtrans, Bartolo:2016ami, Auclair:2019wcv, LISACosmologyWorkingGroup:2022jok,Blanco-Pillado:2024aca,LISACosmologyWorkingGroup:2024hsc,LISACosmologyWorkingGroup:2025vdz}.
\emph{\color{red}{Even in the absence of a detection, strong upper limits on a SGWB of primordial origin will have far-reaching consequences, allowing to constrain many well-studied scenarios that demand observational tests since decades.}}
%ruling out cosmology and particle physics hypotheses that are currently under investigation and theoretical development.   I DO NOT KNOW HOW TO WRITE IT, BUT THE GOAL WOULD TO SAY THAT WE RULE OUT THEORETICAL HYPOTHESES THAT WE TRY TO FALSIFY SINCE DECADES. SO, IT IS IMPORTANT TO DO IT

Detecting {cosmological} SGWBs is challenging. 
Primordial SGWB sources are based on, so far untested, physics beyond the Standard Model (BSM): it is therefore impossible to make definitive  predictions about the signal presence and about its 
characteristics, such as its spectral shape. 
If the SGWB is strong, as in many scientifically motivated examples such as those shown in \cref{fig:allSGWB}, detection could be convincing, leading to a groundbreaking discovery. 
If it is weak, it can be masked by astrophysical foregrounds (some of which are also shown in \cref{fig:allSGWB}).
Furthermore, both types of signals are stochastic and therefore appear as additional (i.e.~not instrumental) noises in the detector data stream:  
distinguishing them from the instrument noise will be challenging, without prior knowledge on their spectral features.
This difficulty can be mitigated for the foregrounds whose amplitude and spectral shape is predicted from observational knowledge of sources population. 
On the other hand, 
the cosmological SGWB detection requires {cutting-edge} data analysis {techniques} to distinguish it from the instrumental noise, currently under development  \cite{Muratore:2021uqj,Muratore:2022nbh,Muratore:2023gxh,Hartwig:2023pft,Alvey:2024uoc,Pozzoli:2023lgz}. 
If the detection succeeds, and both the SGWB amplitude and spectral shape can be reconstructed with sufficient accuracy, the SGWB origin can be inferred.
The measured SGWB parameters can then be translated into a region in the underlying theory’s parameter space, pinning down the discovery space. 
\emph{\color{red}{Precision both in the measurement and in the theoretical prediction is crucial to breaking the degeneracy in the parameter space.}}

Among the many proposals leading to observable cosmological {SGWBs}, we identify three
state-of-the-art science drivers: first order phase transitions, cosmic strings, and inflation.
The constraining power of LISA for these sources has been recently 
analysed in four papers of the LISA Cosmology Working Group \cite{Caprini:2024hue,Blanco-Pillado:2024aca,LISACosmologyWorkingGroup:2024hsc,LISACosmologyWorkingGroup:2025vdz}.

\begin{figure}
    \centering
    \includegraphics[height=0.35\textwidth]{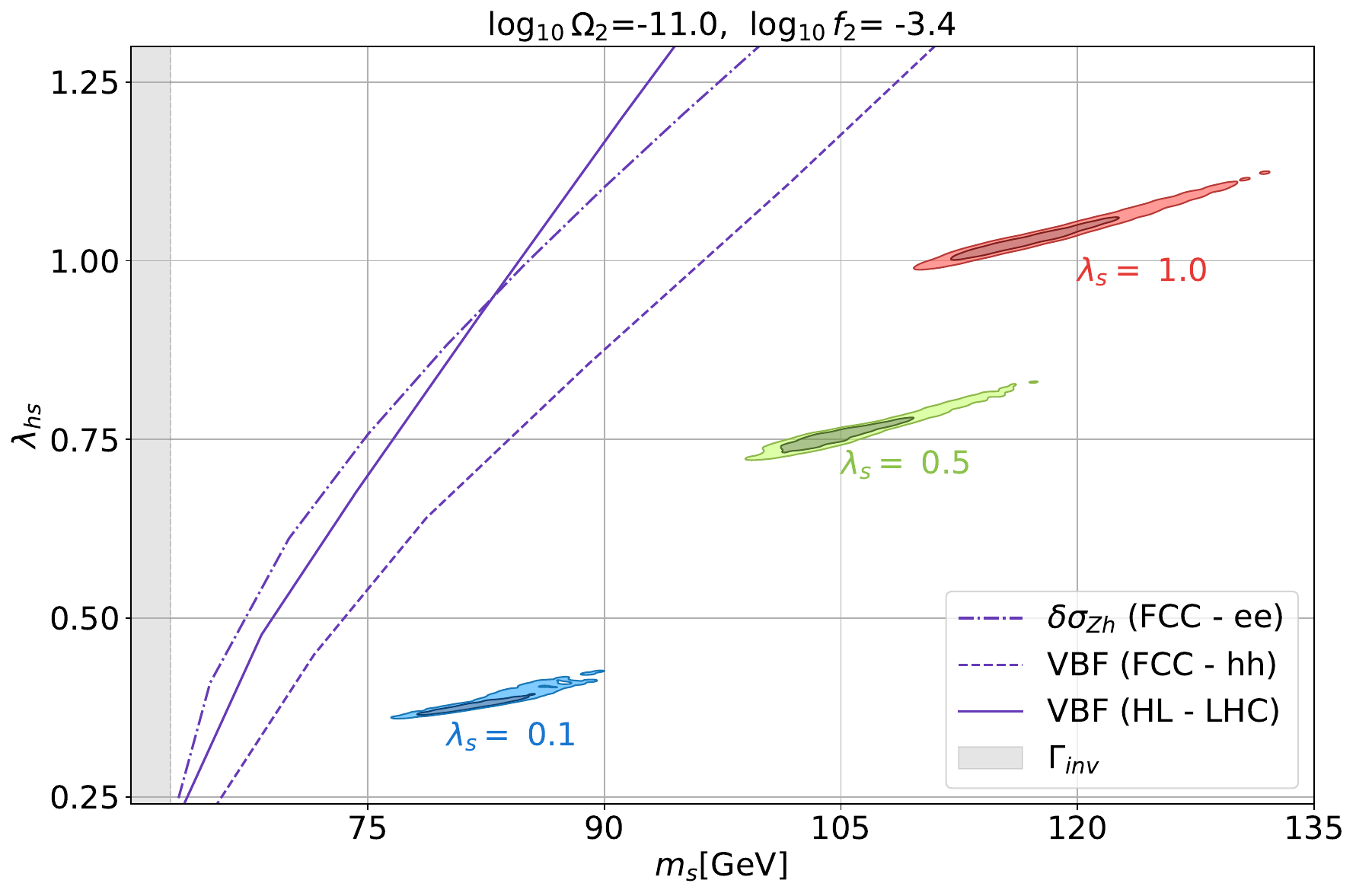} \hspace{2mm}
    \includegraphics[height=0.35\textwidth]{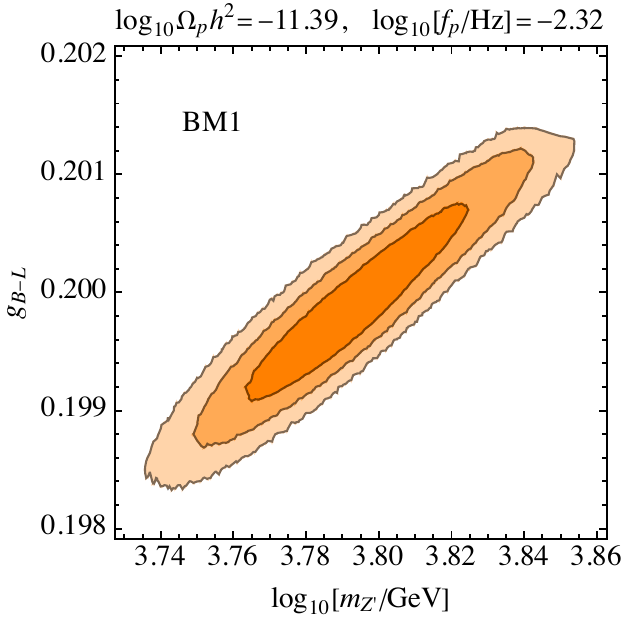}
    \caption{\footnotesize Taken from Ref.~\cite{Caprini:2024hue}. \emph{Left panel}: a FOPT in the context of the $\mathbb{Z}_2$-singlet extension of the SM can lead to a SGWB with peak amplitude and frequency detectable by LISA (given in the top legend). The LISA posterior on the SGWB parameters can in turn be translated into posteriors on the model parameters, namely~the singlet mass $m_{\rm s}$ and higgs-singlet coupling $\lambda_{\rm hs}$: these are represented by the coloured regions, for fixed values of the singlet self-coupling. The experimental sensitivities to Higgs associated production and vector boson fusion of HL-LHC and FCC are shown as violet lines, according to the legend.
    \emph{Right panel:} similarly for the scale-invariant $U(1)_{\rm B-L}$ SM extension, the detection of the SGWB by LISA can be mapped into constraints on the gauge coupling $g_{B-L}$ and gauge boson mass $m_{Z'}$. The shaded regions show the \SI{68}{\%}, \SI{95}{\%}, and \SI{99.7}{\%} confidence level regions provided by the {LISA} measurement, for fixed SGWB peak amplitude and frequency, given in the top legend.}
    \label{fig:FOPT}
\end{figure}

\vspace{0.5em}
$\bullet$~\textbf{{SGWB} from first order phase transitions (FOPT)}: 
well-motivated BSM scenarios predict FOPTs at the electroweak scale and beyond, which can produce SGWBs peaking in the LISA band \cite{Caprini:2015zlo,coswg_PE_phtrans,Caprini:2024hue}.
The SGWB originates from the collision of true vacuum bubbles and subsequent cosmic fluid motion, including sound waves and magnetohydrodynamic turbulence.
Typically, the SGWB peaks at a frequency corresponding to the bubble size at collision, and decays as power laws on both sides.
Precise signal reconstruction allows inferring the thermodynamic parameters of the GW source, related to the FOPT parameters: fractional gradient/kinetic energy density (FOPT strength), bubble size, bubble wall velocity, and FOPT temperature.
\emph{\color{red}{The thermodynamic parameters can be mapped to those of the particle physics model underlying the FOPT, synergizing with current and future particle physics experiments.}}
As an example, Ref.~\cite{Caprini:2024hue} forecast LISA sensitivity to SGWBs from FOPTs, focusing on reconstructing two popular BSM scenarios leading to FOPTs: the {SM} $\mathbb{Z}_2$-singlet extension and the {SM} scale-invariant $U(1)_{\rm B-L}$ extension. Results and collider complementarity are shown in \cref{fig:FOPT}.
From a SGWB with peak amplitude and frequency fixed to values guaranteeing detection, in the former model LISA can constrain the singlet mass and the Higgs-singlet coupling with an accuracy of about 10\% (for a known singlet quartic coupling). 
In the latter model, from the detection of a SGWB with similar amplitude, {LISA} can constrain the new gauge coupling to within 1\%, for values of the boson mass $m_{Z'}$ still allowed by collider searches. 

The fact that dark matter has so far evaded any attempt of detection other than gravitationally, has prompted the idea of dark sectors:  populated by multiple new states, including dark matter candidates, they only interact with the SM through a portal. 
Dark sectors are characterised by new forces and symmetries, and can therefore host strong FOPTs \cite{Schwaller:2015tja,Fairbairn:2019xog,Jaeckel:2016jlh}. 
The detection of the associated SGWB would constitute a valuable channel to grant observational evidence for the existence of these, otherwise elusive, sectors \cite{Aoki:2017aws,Breitbach:2018ddu,Helmboldt:2019pan,Croon:2018erz}. 
LISA is well positioned to detect SGWBs from dark FOPTs occurring at temperatures around and beyond the TeV scale, whose dynamics, however, can be completely disconnected from the electroweak sector of the SM \cite{Bringmann:2023iuz,Kierkla:2022odc,Baldes:2018emh}.

\vspace{0.5em}
$\bullet$~\textbf{{{SGWB} from cosmic strings (CS)}}: 
CSs, topological defects that may form in symmetry-breaking PTs, are powerful GW sources \cite{Kibble:1984hp,Damour:2001bk}. 
The traditional example is associated to a local $U(1)$ symmetry breaking in the Abelian-Higgs model, but depending on the theory underlying their formation, CSs can be of different nature: more complex symmetry breaking patterns can lead, for example, to string with internal structure (superconducting or metastable strings); they can also occur as fundamental objects of string theory (cosmic super-strings). 

The GW signal depends on the CS properties. For instance, strings associated with a global symmetry breaking may radiate mainly Goldstone bosons, while CSs described by the Nambu-Goto action copiously produce GWs. 
CSs form a universe-filling network of super-horizon sized strings and smaller closed loops continuously sourcing GWs during its evolution. 
This leads to a SGWB with nearly flat spectrum over a wide frequency range, and amplitude proportional to the square of the dimensionless string tension $G\mu$, which is related to the energy scale $\eta$ of the PT by $G\mu\sim 10^{-6}[\eta/(10^{16} {\rm GeV})]^2$ \cite{Auclair:2019wcv}. Crucially, PTs at high energy, such as the GUT scale, correspond to large SGWBs, providing a unique opportunity to probe high-scale BSM physics beyond collider capabilities.

In this context, LISA will grant significant advancement. {For a given CS model,}
LVK constrained the string tension to $G\mu \lesssim 9.6 \times 10^{-9}$ \cite{LIGOScientific:2021nrg}, while the best-fitting value for the EPTA DR2 data set is $G\mu \simeq 7.9 \times 10^{-11}$ \cite{2023arXiv230616227A}.
Accounting for astrophysical foregrounds, LISA will be able to probe the SGWB from CSs with tension down to $G\mu\sim 10^{-16}$, 
with a precision within one order of magnitude \cite{Auclair:2019wcv,Blanco-Pillado:2024aca}.
\emph{\color{red}{This tension corresponds to the symmetry breaking scale $\eta \sim 10^{11}\,{\rm GeV}$: 
LISA will thus probe a huge portion of the CS parameter space that would remain otherwise unexplored, five orders of magnitude below the GUT scale currently probed by LVK and PTA}}.
These figures depend on the CS model. Even in the Nambu-Goto scenario, the GW signal varies based on loop density, size, and GW emission. However, LISA improvement over LVK and PTA remains groundbreaking. \emph{\color{red}{Conclusive evidence of a CSs network would revolutionise BSM physics, even with the current CS model uncertainty.}}

\vspace{0.5em}
$\bullet$~\textbf{{{SGWB} from inflation}}: 
{inflation \cite{Guth:1980zm,Starobinsky:1980te,Linde:1981mu,Albrecht:1982wi} addresses cosmological issues and explains primordial fluctuations consistent with observations.
Yet, constraining inflation remains a key challenge in modern cosmology.}
Thanks to its sensitivity to mHz GWs, \emph{\color{red}{LISA will
probe inflation in a new observational window, corresponding to comoving wave-numbers $k \sim [10^{10}, 10^{14}] \,{\rm Mpc}^{-1}$, ten to fifteen orders of magnitude larger than those probed by CMB and Large-Scale Structure}} \cite{Lasky:2015lej,LISACosmologyWorkingGroup:2022jok}.

The tensor metric fluctuations are observationally bounded at CMB scales (femto-Hz) \cite{BICEP:2021xfz}.  Assuming a red-tilted slow-roll spectrum, this extrapolates to $h^2 \Omega_{\rm GW} \lesssim 10^{-16}$ at LISA frequencies, well below its sensitivity. 
However, several well motivated inflationary scenarios go beyond the simplest setup, and can result in significantly greater SGWBs at LISA frequencies \cite{Bartolo:2016ami,LISACosmologyWorkingGroup:2022jok,LISACosmologyWorkingGroup:2024hsc}, offering the unique opportunity to obtain direct observational information on late inflationary dynamics.

An inflationary SGWB detectable by LISA can be sourced either directly \cite{Starobinsky:1979ty}, or at second order from scalar perturbations \cite{Tomita:1975kj, Matarrese:1992rp,Matarrese:1993zf, Acquaviva:2002ud,Ananda:2006af}. 
Direct production arises, for example, in axion inflation, where a pseudo-scalar inflaton drives the production of gauge field modes that in turn source GWs, enhancing the SGWB in the LISA band~\cite{Barnaby:2010vf,Sorbo:2011rz}. 
The SGWB amplitude is exponentially sensitive to a parameter inversely proportional to the axion decay constant, while its spectral tilt  is shaped by first and second slow-roll parameters evaluated when the relevant frequencies exit the Hubble sphere. 
Ref.~\cite{LISACosmologyWorkingGroup:2024hsc} finds that LISA can constrain axion inflation parameters, including the first slow-roll parameter $\epsilon_*$, to sub-1\% precision. 
Other models providing directly enhanced SGWBs testable by LISA include the breaking of space diffeomorphisms \cite{Endlich:2012pz,Ricciardone:2016lym,Bartolo:2015qvr}, which can give
rise to a massive graviton during inflation, tilting the SGWB spectrum; or when a large
number of scalar particles are produced dynamically, with momenta peaked around a narrow range of scales, and the energy-momentum tensor of the
produced particles sources the SGWB \cite{Fumagalli:2021mpc}. 

Scalar perturbations generate tensor perturbations at second order in perturbation theory \cite{Tomita:1975kj, Matarrese:1992rp,Matarrese:1993zf, Acquaviva:2002ud,Ananda:2006af}.
While small at CMB scales, scalar-induced SGWBs can be significant at smaller scales, if scalar perturbations are amplified there. 
\emph{\color{red}{A detection or upper bound on the SGWB from scalar perturbations can be translated into a revolutionary probe of the scalar amplitude  $\mathcal{A}_s$ beyond CMB scales}}. 
Enhanced scalar perturbations naturally arise in single-field inflationary scenarios with features in the potential, like those leading to an ultra-slow-roll phase; in multi-field inflationary setups, featuring
a slow-turn in the field space together with a tachyonic instability of isocurvature perturbations, or a short period of non-attractor evolution; if localised sharp features are present in the inflationary potential (steps, sharp turns in the inflationary trajectory, ...). See \cite{LISACosmologyWorkingGroup:2025vdz} for an overview.
The only current bound at LISA scales, derived from CMB constraints on extra relativistic species, is $\mathcal{A}_s\leq 0.69$. Refs.~\cite{LISACosmologyWorkingGroup:2024hsc,LISACosmologyWorkingGroup:2025vdz} find that LISA can improve this to $\mathcal{A}_s\leq 10^{-3.5}$ (99.7\% CL) for models featuring double-peaked or broken power-law SGWBs (covering e.g.~ultra-slow-roll phases or multi-filed hybrid inflation). 

Intriguingly, the formation of primordial black holes (PBHs) require large perturbations that inevitably seed sizeable SGWBs at second order.
By detecting or setting 
upper bounds on scalar-induced SGWBs in the mHz band, LISA probes not only the model of inflation, but also the only mass window in which PBHs can constitute the totality of the dark matter, namely asteroidal-mass $[10^{-16},10^{-10}]~M_\odot$ \cite{Bartolo:2018evs,Carr:2020gox}.
\emph{\color{red}{LISA will therefore discover, or rule out the idea, that the totality of dark matter is made of PBHs formed from the collapse of moderately non-Gaussian
curvature fluctuations}} \cite{LISACosmologyWorkingGroup:2023njw}. 
Ref.~\cite{LISACosmologyWorkingGroup:2024hsc} finds that LISA can measure the mean PBH mass with sub-percent accuracy and constrain the PBH dark matter fraction to better than 10\%.

\subsubsection*{Measure, or set upper limits on, the spectral shape of the cosmological {SGWB}}
\label{sec:CosmoSGWB}

We present two proof-of-concept analyses (extracted from \cite{colpi2024lisadefinitionstudyreport}) illustrating that \emph{i)} a cosmological {SGWB} can be identified in the LISA data, and \emph{ii)} its spectral shape can be reconstructed, under two simplifying assumptions: an equal-arm detector (exploiting the signal-orthogonal $T$-channel) and a parametrised model for the instrument noise. 
Because of their inherent uncertainty, cosmological {SGWBs} must be searched 
for by minimising a priori assumptions on their spectrum, while remaining as agnostic as possible about the instrument
noise. Unlike in the science drivers examples above, here the signal inference is not template-based: we apply agnostic searches, aimed at reconstructing the signal spectral shape. 
Our benchmark (displayed in \cref{fig:allSGWB}) is a {SGWB} generated within the SM singlet extension  \cite{Caprini:2024hue}, for which we assume the spectral shape
$ h^2\Omega_{\rm PT}(f) = h^2\Omega_p \left(f/f_p\right)^3 \left\{7/ \left[4 + 3\left(f/f_p\right)^2\right]\right\}^{7/2}$ \cite{coswg_PE_phtrans}, further setting $ h^2\Omega_p= \num{e-10}$ and $f_p=\qty{2e-4}{\Hz}$. \emph{\color{red}{Determining the {SGWB} spectral shape is the simplest resource available to assess its origin.
LISA can identify the presence, and reconstruct basic features in the spectral shape, of physically motivated cosmological {SGWBs}.}}

\begin{figure}
\centering
\includegraphics[width=0.35\textwidth]{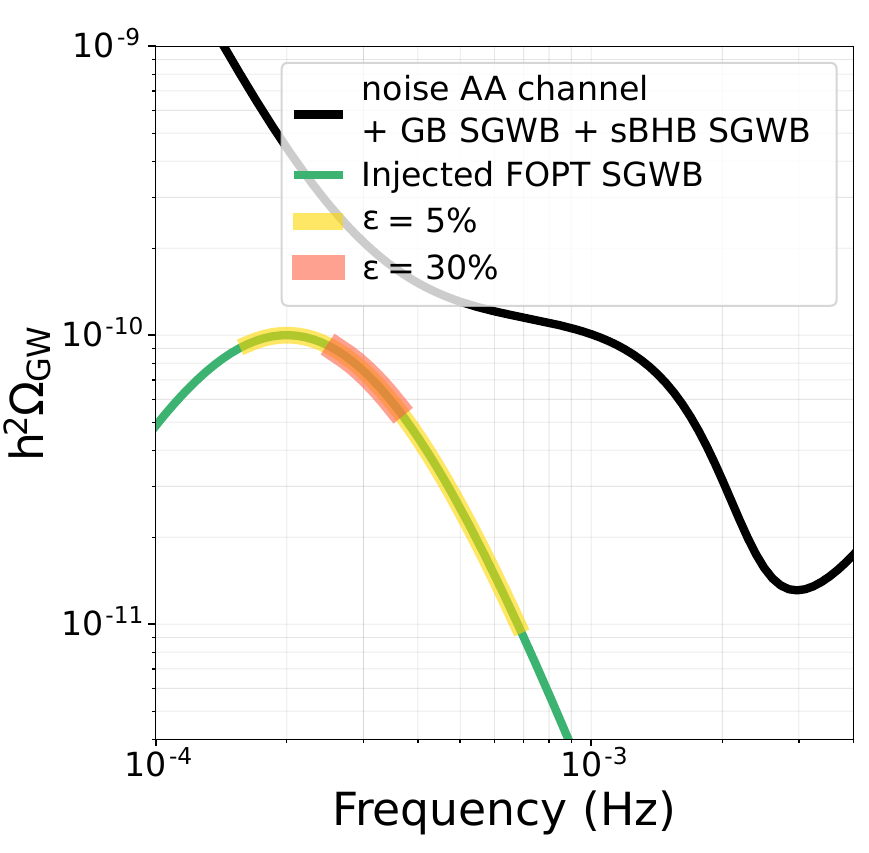}
\includegraphics[width=0.64\textwidth]{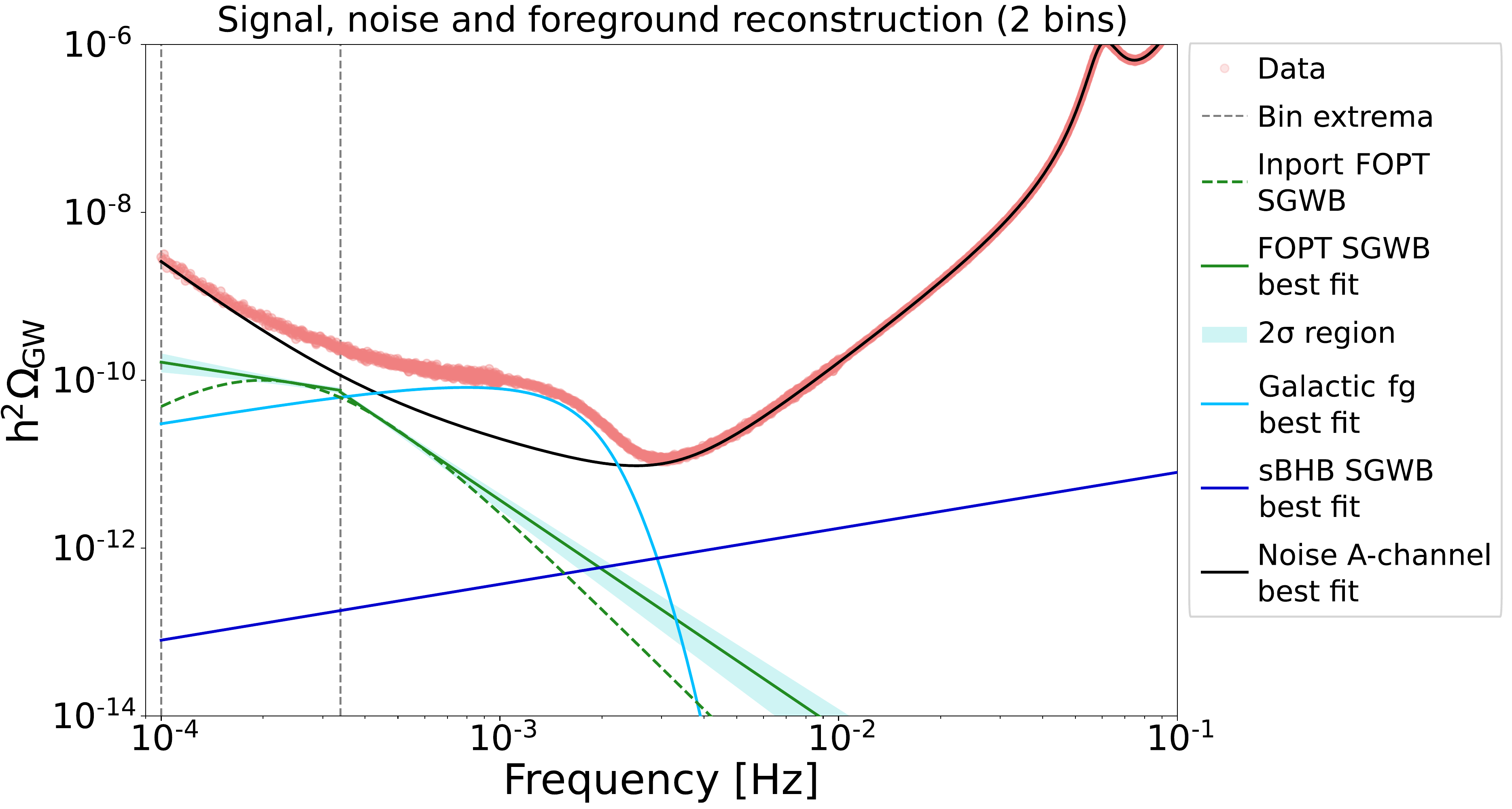}
\caption{\footnotesize Taken from \cite{colpi2024lisadefinitionstudyreport}. \emph{Left panel}:
the identification of a cosmological {SGWB}. The presence of the injected {FOPT} {SGWB} (green line) can be identified in the simulated LISA data imposing a hard prior on the instrument noise, for two tested values of the relative uncertainty on the noise + foregrounds:  $\varepsilon={5}{\%}$ and ${30}{\%}$. The frequency region over which the signal is identified (Bayes factor $\log \mathcal{B}_{\rm FOPT}>50$) is shown in orange for $\varepsilon={30}{\%}$ and yellow for $\varepsilon={5}{\%}$.
 \emph{Right panel}: 
The outcome of a signal-reconstructing procedure based on searching for the {FOPT} {SGWB} (green, dashed line: injected signal) as a series of power laws in frequency bins.
  The SGWBinner code has been run on simulated data (red dots) containing \emph{i)} the {FOPT} {SGWB}, \emph{ii)} the {sBHBs} foreground (dark blue) and Galactic foreground (light blue), \emph{iii)} the instrument noise (black). The code has iteratively merged several initial frequency bins in two final ones. 
    The  presence of a break in the {FOPT} {SGWB} can be reconstructed piece-wise, with two power laws (green, solid).
   The 2$\sigma$ error on the reconstructed signals and noise are shown as shaded areas.}
\label{fig:FoM2_12}
\end{figure}

\vspace{0.5em}
$\bullet$~\textbf{Identification of the SGWB}: Following the procedure of \cite{Karnesis:2019mph}, the presence of our benchmark signal can be identified in simulated LISA data containing \emph{i)}
the signal itself, \emph{ii)} the Galactic and sBHBs foregrounds, \emph{iii)} the instrument noise.  
The LISA frequency range is binned, and the presence of the cosmological {SGWB} is identified bin by bin, assuming a hard prior on the other components. Two values for the relative uncertainty of the noise 
+ foregrounds are tested: $\varepsilon=5\%$ and $\varepsilon=30\%$.  
If the Bayes factor between the model with noise + foregrounds + 
{SGWB} and the model with only noise + foregrounds satisfies $\log \mathcal{B}_{\rm FOPT}>50$ in at least one
bin, we consider that the first model is preferred and the presence of the SGWB can be identified in that bin.  
The result is summarised in the left panel of
\cref{fig:FoM2_12}: the benchmark {SGWB} (green curve) can be identified in the data over a sizeable range of frequencies, highlighted in yellow for $\varepsilon={5}{\%}$ and in orange for $\varepsilon={30}{\%}$.

\vspace{0.5em}
$\bullet$~\textbf{Reconstruction of the SGWB}: The SGWBinner code \cite{Caprini:2019pxz,Flauger:2020qyi} aims at piece-wise reconstructing the {SGWB} spectral shape by searching, in a series of frequency intervals, for disconnected power-law signals. 
Simultaneously, a
template-based search is performed for the instrument noise and the foregrounds. 
%The noise template has two parameters: A (mass acceleration
%noise) and P (optical metrology system noise) \cite{Flauger:2020qyi}.  The prior for the noise parameters is established using the
%$T$-channel.  
The posterior for all components is minimised independently in each bin, over the full set of parameters: the two noise parameters $(A,P)$ (mass acceleration
and optical metrology system noises), the two foregrounds amplitudes $(A_{\rm GB},A_{\rm sBHB})$ (their spectral shape is assumed), and two {SGWB} parameters per
bin $i$: the spectral index $n_i$, and the amplitude $\Omega_i$ at pivot frequency (given by the geometrical mean of the bin $i$). 
For all pairs of neighbouring bins, the code iteratively checks whether merging them is statistically favoured.  
The
result on the benchmark SGWB is shown in the right panel of \cref{fig:FoM2_12}: the merging procedure gives two final bins, allowing to recognise the presence of a break in the {SGWB} spectral shape (solid green lines). 
Ref.~\cite{colpi2024lisadefinitionstudyreport} also reports the {SNR} and the marginalised
$2\sigma$ errors on all the model parameters.

\vspace{0.5em}

Through the SGWB detection, LISA can test early universe scenarios. 
In many models, the SGWB parameters can be constrained with exquisite accuracy, provided the instrument noise and the astrophysical foregrounds are well characterised, and the residuals from the deterministic source subtraction are limited. 
In order to  enforce the opportunity of SGWB detection at the forecasted statistical errors, in the next years before the LISA launch, \emph{\color{red}{the community should focus on {i)} reducing the theoretical uncertainties, identifying potential smoking-gun signatures of specific scenarios and addressing degeneracies between models, and {ii)} developing data analysis techniques aimed at optimising how to distinguish the noise, the foreground, and the SGWB, taking into account the respective uncertainties. 
This would allow to fully exploit the exquisite reconstruction precision that LISA offers, fully enabling its discovery potential.}}

\subsection{SO8: Search for GW bursts and unforeseen sources}

When opening a new observational band, the exciting prospect is discovering unexpected sources.
The mission design aims at minimising non-stationary noises and artefacts, with adequate calibration, to facilitate the detection of strong unknown sources. Furthermore, data analysis pipelines will explicitly consider unmodelled sources. 
These can take the form of well-separated-in-time {GW} bursts. One known example of such source-type is the GW emission from CS loops, and in particular from cusps (points on the string momentarily moving at the speed of light) and kinks (discontinuities in the string tangent vector due to strings intersections) on loops \cite{LISACosmologyWorkingGroup:2022jok}. 
These features emit a beam of polarised GW radiation, which depends on the loops redshift and length, and whose beaming angle is larger at lower frequency~\cite{Damour:2000wa,Damour:2001bk}, enhancing the probability of burst detection at LISA with respect to LVK. 
The SNR of a burst in LISA depends on the GW amplitude, its polarisation, and the sky localisation of the event. 
The expected burst rate further depends on the loop distribution function \cite{LIGOScientific:2021nrg}, and has been determined for LISA in \cite{auclair2023cosmic}, as a function of $G\mu$ and setting $\mathrm{SNR}>20$ as the threshold for single-burst detection. 
The non-detection of CS bursts in LISA would constrain the string tension down to $G\mu \lesssim \num{e-11}$. Alternatively, if $G\mu \approx \num{5e-11}$ (compatible with the {PTA} signal), LISA would observe between $4-30$ CS bursts per year \cite{colpi2024lisadefinitionstudyreport}.

\section{Conclusion}

By opening the observational window of mHz GWs, LISA grants access to new classes of GW sources with high scientific potential, paving the way for groundbreaking discoveries. 
LISA puts Europe in a world-leading position for future GW physics. 
It is paramount to ensure the maximal science return from the mission, in order to best serve the broad, interdisciplinary community of scientists interested in the LISA science case, spanning astrophysics, fundamental physics, cosmology. 

We have identified four among LISA SOs in close relation with the focus of the European Strategy for Particle Physics. 
LISA provides tests of physical phenomena in its own peculiar way, often complementary to more targeted probes such as cosmological observatories and particle colliders.
In particular, we have reported that LISA has the capability to \emph{i)} test the theory of gravitation, in both the strong and weak field limit; \emph{ii)} complement, via a new messenger, cosmological tests of the expansion of the universe, even at high redshift; \emph{iii)} unveil the physics of the early universe, and therefore probe high energy physics, through the detection of SGWBs of cosmological origin. 
Therefore, LISA can address fundamental questions such as the nature of black holes and of the gravitational radiation; the occurrence of spontaneous breakings of the fundamental
symmetries of the theory of particle interactions; the generation of the matter-antimatter
asymmetry of the universe; the production of dark matter candidates of diverse nature; the presence and characteristics of dark energy.

Furthermore, LISA will operate together with PTAs and next-generation Earth-based detectors, and combining information at different frequency bands can also be transformative for GW science.

%\href{https://www.overleaf.com/learn/how-to/Including_images_on_Overleaf}{including images on Overleaf}.

%\begin{figure}
%\centering
%\includegraphics[width=0.25\linewidth]{frog.jpg}
%\caption{\label{fig:frog}This frog was uploaded via the file-tree menu.}
%\end{figure}

%\begin{table}
%\centering
%\begin{tabular}{l|r}
%Item & Quantity \\\hline
%Widgets & 42 \\
%Gadgets & 13
%\end{tabular}
%\caption{\label{tab:widgets}An example table.}
%\end{table}

\newpage

\noindent \textbf{\large{Acknowledgments:}} 
R.B.~acknowledges financial support provided by FCT – Fundação para a Ciência e a Tecnologia, I.P., through the ERC-Portugal program Project ``GravNewFields''.
A. H. is supported by grant PD-034-2023 co-financed by the
Govern Balear and the European Social Fund Plus (ESF+) 2021-2027. 
This work was supported by the Universitat de les Illes Balears (UIB); the Spanish Agencia Estatal de Investigación grants PID2022-138626NB-I00, RED2022-134204-E, RED2022-134411-T, funded by MICIU/AEI/10.13039/501100011033 and the ERDF/EU; and the Comunitat Autònoma de les Illes Balears through the Conselleria d'Educació i Universitats with funds from the European Union - NextGenerationEU/PRTR-C17.I1 (SINCO2022/6719) and from the European Union - European Regional Development Fund (ERDF) (SINCO2022/18146).
N.T.~acknowledges support form the French space agency CNES in the framework of LISA.

\bibliographystyle{JHEP}
\bibliography{biblio.bib}

\end{document}